\documentclass[12pt,preprint]{aastex}
\usepackage{epsfig}
\usepackage{graphics}
\usepackage{graphicx}

\shorttitle{SVS~13}
\shortauthors{Hodapp}

\begin{document}

\title{The Launch Region of the SVS~13 Outflow and Jet\\}

\author{Klaus~W. Hodapp\altaffilmark{1},
Rolf~Chini\altaffilmark{2}$^,$\altaffilmark{3} }

\altaffiltext{1}{
Institute for Astronomy, University of Hawaii,
640 N. Aohoku Place, Hilo, HI 96720, USA
\\email: {\em hodapp@ifa.hawaii.edu} }

\altaffiltext{2}{
Astronomisches Institut, Ruhr-Universit{\"a}t Bochum, 
Universit{\"a}tsstra{\ss}e 150, D-44801 Bochum, Germany
\\email: {\em rolf.chini@astro.ruhr-uni-bochum.de}
}

\altaffiltext{2}{
Instituto de Astronomia, 
Universidad Catolica del Norte,
Avenida Angamos 0610, Antofagasta, Chile
}

\begin{abstract} 

We present the results of Keck Telescope laser adaptive optics integral field
spectroscopy with OSIRIS of the innermost regions of the NGC~1333 SVS~13 outflow that
forms the system of Herbig-Haro objects 7-11. We find a bright 0$\farcs$2 long microjet traced by
the emission of shock-excited [FeII]. Beyond the extent of this jet, we find a series of
bubbles and fragments of bubbles that are traced in the lower excitation
H$_2$~1--0~S(1) line.
While the most recent outflow activity is directed almost precisely (PA~$\approx$~145$^{\circ}$) to the
south-east of SVS~13, there is clear indication that prior bubble ejections were pointed in 
different directions.
Within these variations, a clear connection of the newly observed bubble ejection events 
to the well-known, poorly collimated HH 7-11 system of Herbig-Haro objects is established. 
Astrometry of the youngest of the expanding shock fronts at 3 epochs covering a time span of over two years
gives kinematic ages for two of these.
The kinematic age of the youngest bubble is slightly older than the historically observed
last photometric outburst of SVS~13 in 1990, consistent with that event launching the bubble and some
deceleration of its expansion.
A re-evaluation of historic infrared photometry and new data show that SVS~13 has not 
yet returned to its brightness before that outburst and thus shows a
behavior similar to FUor outbursts, albeit with a smaller amplitude.
We postulate that the 
creation of a series of bubbles and the changes in outflow direction are indicative
of a precessing disk and accretion events triggered by a repetitive phenomenon 
possibly linked to the orbit of a close binary companion. 
However, our high-resolution images in the H and K bands do not
directly detect any companion object.
We have tried, but failed to detect, the kinematic signature of rotation 
of the microjet in the [FeII] emission line at 1.644~$\mu$m.

\end{abstract}

\keywords{
infrared: stars ---
ISM: Herbig-Haro objects ---
ISM: jets and outflows ---
stars: formation --- 
stars: individual (SVS~13) ---
stars: variables: other 
}

\section{INTRODUCTION}
The accretion of matter onto a forming star is inextricably associated
with mass outflow. This manifests itself as well-collimated jets typically
observed in higher excitation shock-excited forbidden lines such as [SII] or [FeII], 
or lower excitation emission
of shock-exited H$_2$. Larger-scale and older outflows are typically detected
in CO emission of entrained ambient material. While the precise mechanism of jet launching is 
actively being debated, steady progress has been made in observational
studies of the launch regions of jets. Such studies require the highest
possible spatial resolutions, since all theories postulate that jets
are launched from the inner regions of the protostellar disks on spatial scales
of a few AU, or from the protostar's magnetosphere, on scales of a stellar
diameter.

At present, the highest spatial resolutions for studies of the jet launch
regions are achieved at
optical wavelengths with the Hubble Space Telescope (HST), and at near-infrared
wavelengths with adaptive optics on large ground-based telescopes.
The near-infrared techniques have the advantage of being able to better
penetrate dust extinction, so that the jet launch regions of more
deeply embedded, generally younger stars can be observed.
However, near-infrared laser-guide-star adaptive optics observations
today are still
limited by the requirement to have a fairy bright, optically visible
(R $\lessapprox$ 16) tip-tilt reference star close to the object, or that
the object itself can serve this purpose.
As a consequence, even at near-infrared wavelengths, the best candidates for high spatial
resolution studies are objects
near the end of their Class I phase or early in the classical T Tauri star
phase because in nearby molecular (dark) clouds, 
the only available optical tip-tilt reference star is often the young
star itself.

In this paper we present detailed adaptive-optics corrected integral field 
spectroscopy of NGC~1333 SVS~13, the driving source of the famous chain
of Herbig-Haro objects HH~7-11 \citep{Herbig1983}. Other commonly used names for
SVS~13 are V512 Per and 2MASS J03290375+3116039. The SVS~13 outflow appears relatively poorly
collimated and is comprised of a number of individual shock fronts, as can
be seen in Fig.~1 that presents a progression from seeing-limited to diffraction-limited
images of the SVS~13 outflow.
There is an anti-parallel counter outflow visible, but it is displaced
from the axis of the HH~7-11 outflow. For a general overview of NGC~1333 and
a review of the literature on the SVS~13 subcluster, the reader is referred
to the review article by \citet{Walawender2008}.
The Two Micron All Sky Survey (2MASS) position of SVS~13 is
$3^{h}~29^{m}~03^{s}.759~+31^{\circ}~16^{\prime}~03\farcs99~(J2000)$ at epoch 1999 Nov. 26
with an accuracy
of $\pm~0\farcs10$
\citep{Skrutskie2006}.
This position lies between the VLA configuration A radio position 
of VLA 4B at 3.6 cm reported by \citet{Anglada2000} at
$3^{h}~29^{m}~03^{s}.74~+31^{\circ}~16^{\prime}~04\farcs15~(J2000)$ with an effective
epoch of 1997.9 and estimated errors of $\pm~0\farcs05$ and
the 7 mm VLA configuration B position of
the same radio source, which \citet{Anglada2004} give as 
$3^{h}~29^{m}~03^{s}.759~+31^{\circ}~16^{\prime}~03\farcs94~(J2000)$
with an epoch of 2001 May 4.
The epoch of the 2MASS observations is between the epochs of the VLA observations,
and the reported near-infrared coordinates lie between those VLA coordinates. Within the errors
of $\pm~0\farcs05$ for the VLA data and
of $\pm~0\farcs10$ for 2MASS, and a tentatively indicated small proper motion of VLA~4B, 
these data show that the near-infrared source SVS~13
is identical to the VLA mm and cm-wavelength source VLA~4B.
It should be noted that \citet{Anglada2000} and \citet{Anglada2004},
on the basis of older optical astrometry,
had originally identified SVS~13 with a different cm-radio source: VLA~4A.

\begin{figure}
\begin{center}
\figurenum{1}
\includegraphics[scale=0.9,angle=0]{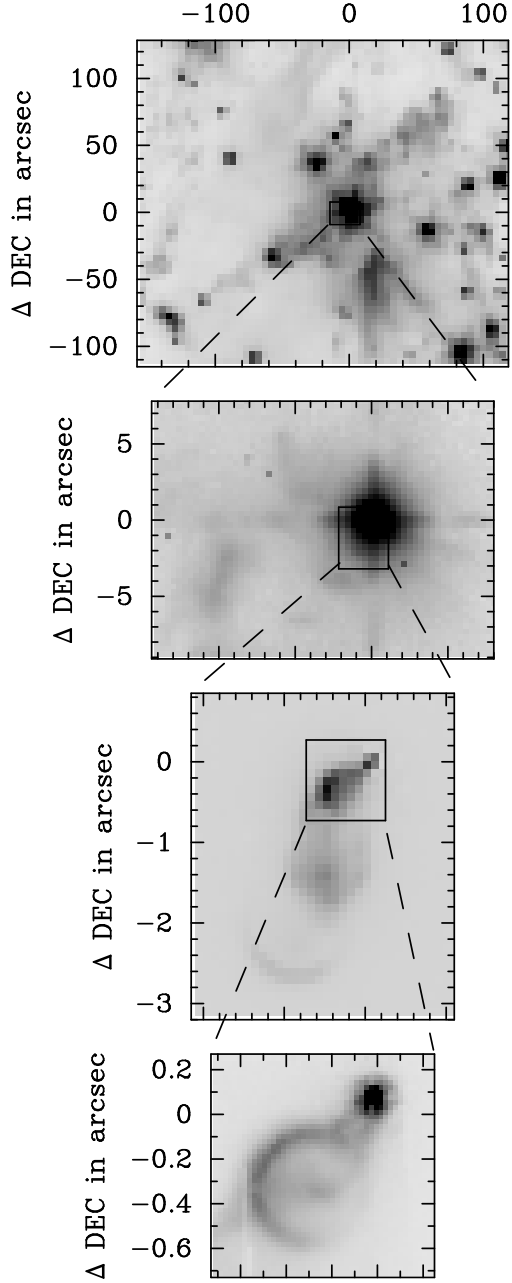}
\caption{
The SVS~13 outflow at four different spatial scales: The top image is a 
Spitzer channel 2 (4.5~$\mu$m) image based on the same original data as those used by
\citet{Raga2013} for their proper motion study.
The image shows the full
extent of the shocked molecular hydrogen emission.
The second panel from the top is a small portion of a 
H$_2$ S(1) 2.122~$\mu$m
image obtained in 1996 with the QUIRC camera at the UH~2.2~m telescope that
shows the emission features closest to SVS~13.
The third panel
shows the integrated H$_2$~S(1) line intensity observed 
with OSIRIS on Keck~II in 2011 with the 100 mas spaxel scale, while the fourth
panel was taken with OSIRIS on Keck~I in 2012 with the 20~mas spaxel scale.
}
\end{center}
\end{figure}

We use the distance to NGC~1333 SVS~13 determined by \citet{Hirota2008} from 
VLBI parallax measurements of masers associated with this object: 235~$\pm$~18~pc.
The study of the dynamics of the NGC~1333 region by \citet{Warin1996} has determined the systemic
velocity of the gas near SVS~13 to be 8~km~$s^{-1}$ relative to the local
standard of rest. Throughout this paper, we will give velocities relative to
this systemic velocity of the NGC~1333 molecular core that SVS~13 is embedded in.

Young, still accreting stars often experience eruptive changes in brightness
that traditionally get classified into either FU Orionis (FUor) or EX Lupi (EXor) type
outbursts, depending on the duration of the outburst and its spectrum.  
The first review of these phenomena has been given by \citet{Herbig1977}.
FUor outbursts exhibit time scales of decades to centuries and the spectral
characteristics of an optically thick luminous disk with absorption line spectra, 
while the less substantive EXor outbursts have timescales of years, have
been observed to actually return to pre-outburst brightness, and show 
optically thin emission line spectra.
Jets emanating from young stars are often comprised of a series of individual shock fronts that
have been postulated to arise from changes in the jet velocity as a result
of repetitive eruptive events, as described in the review by \citet{Reipurth2001} and references
therein. 
This gives, in principle, a way to study the history
of such eruptive accretion instabilities from the ``fossil'' record in the jet
shock fronts. 
Our target object, SVS~13, 
is certainly in this class of multi-shock outflow sources and
has been observed to undergo
an outburst around 1990, but the light curve does not match either the classical
FUor or EXor curves. The relationship of this last observed outburst to the
structure of the SVS~13 outflow will be studied here.

The proper motions of the shock fronts of the HH 7-11 system associated
with SVS~13 have been studied at optical wavelengths by 
\citet{Herbig1983} and later by \citet{Noriega2001},
and in the infrared in the H$_2$ 1--0 S(1) line by \citet{Noriega2002} using the
Near-Infrared Camera and Multi-Object Spectrometer (NICMOS),
by \citet{Khanzadyan2003} using the United Kingdom Infrared Telescope (UKIRT), and with the Spitzer Space Telescope by \citet{Raga2013}.
These studies all gave proper motions of the shock fronts in the range of 33 mas yr$^{-1}$
that establish a kinematic expansion age of 2100 yrs for the most distant shock front
studied there, the HH~7 bow shock.

In general, the jets seen in optical and near-infrared forbidden lines, mostly in
[SII] and [FeII], offer the most direct view of the material
ejected from an accretion disk or from a protostar's magnetosphere.
There are differences in the details of theoretical models of this
process, e.g., in \citet{Camenzind1990}, \citet{Ferreira1997}, \citet{Konigl2000},
or \citet{Shu2000}, but they all agree on the main point:
Jets originating from the magnetosphere of a rotating star, or from a
rotating disk, and detected in high-excitation shock-excited lines
are expected to carry away excess angular momentum and thereby enable
the mass accretion process.

However, to date, the most convincing detection of a rotational signature of jets from
very young and still deeply embedded objects in SED classes 0 and I have
been obtained with radio interferometric observations of outflows in
CO and SiO emission. These observations typically achieve spatial resolutions
of a few arcseconds but make up for this disadvantage by exquisite velocity
resolution.
Carbon-Monoxide high-velocity emission
is mostly associated with the entrainment of ambient molecular material
into a jet originally emitted at even higher velocity. Any signature of the original
jet rotation is expected to be highly confused at this turbulent interface between
the jet and the ambient material.
Nevertheless, \citet{Launhardt2009} detected the kinematic signature of rotation
in the jet of CB~26 in the CO(2-1) line, very close to the source of the jet. 
\citet{Zapata2010} found some evidence for rotational components in the Ori-S6 outflow using
various higher transitions of CO and SO,
while \citet{Pech2012} recently reported a rotational
signature in the HH~797 outflow in IC~348 using CO(2-1). 
In these cases, the strongest rotational signatures were found at some distance ($\approx1\arcsec$) from
the driving source.
Even farther away from the driving source, 
using SiO data from the VLA, 
\citet{Choi2011} found a rotation signature of the 
NGC~1333 IRAS~4A2 protostellar jet, consistent with the disk 
rotation in that object. The jet could only be observed at distances of more than
5$\arcsec$ from its source, and the rotation signature appears most pronounced at distances of
20$\arcsec$.
These results are far from
universal, however. In a series of papers on HH~211 culminating in \citet{Lee2009} they obtained only a
tentative detection of rotation.  \citet{Codella2007} studied the kinematics of SiO emission
in the HH~212 jet and did not detect a credible rotational signature.
In this paper we have tried to detect the kinematic signature of jet rotation in our [FeII] data 
and are presenting our velocity data, but in the
end failed to detect this effect.

\section{OBSERVATIONS}

\subsection{Keck Adaptive Optics Imaging Spectroscopy}

There are two strong systems of near-infrared emission lines that are commonly used
for the study of protostellar jets: the ro-vibrational lines of H$_2$ and
the forbidden [FeII] lines. As was already discussed specifically in the case of SVS~13 by \citet{Takami2006}, 
the H$_2$ lines, the brightest being the 1--0 S(1) line at
2.122~$\mu$m, trace low-velocity shocks, either internal shocks within the jet, or
the turbulent interface to the ambient medium around the jet. This line is therefore well suited for the
study of internal shocks and entrainment of ambient material by the jet, but is
less suited for a detection of the jet itself.

In contrast, [FeII] traces higher temperatures and excitations and
is usually confined to the densest parts of a jet near its launch region, and to
its terminal shocks against ambient material. The [FeII] lines in the near-infrared are therefore particularly
suited for a study of the kinematics of the jet itself.
Emission from atomic hydrogen, in the near infrared specifically the Brackett series
of lines, is observed in many young stars, and typically
originates in the spatially unresolved accretion disk around the young star itself,
and not in the jet. 

The adaptive optics data reported here were obtained at the Keck~I and II telescopes, using
laser guide star adaptive optics in conjunction with the OH-suppression Infrared Imaging Spectrograph (OSIRIS)
built by \citet{Larkin2006}. 
OSIRIS is a lenslet integral field spectrograph where each spatial element (spaxel) forms an image of
the telescope pupil before being dispersed into a spectrum. This technique separates spatial flux gradients
in each spaxel from any velocity shifts and is therefore very suitable for measuring small shifts in line
centroid over extended objects with spatial scales of the order of a spaxel.
The spectral lines mentioned above can be observed in two setting of OSIRIS. 
In the 3$^{rd}$ grating order and
through the Kn2 filter, the
H$_2$ S(1) line at 2.122~$\mu$m is covered. 
In the 4$^{th}$ grading order and through the Hn3 filter, 
two [FeII] lines and two Brackett-series lines are included in the spectral
bandpass.

The first observations were carried out in the night of 2011 August 21, (UT) (MJD 55794.5345 = 2011.638) with
OSIRIS on the Keck~II telescope, using the relatively coarse 100~mas per spaxel and 50~mas per spaxel scales and the Kn2 and Kbb filters.
A second, higher quality set of data was obtained in the night of 2012 November 4, (UT)
(MJD 56235.4853 = 2012.844) with OSIRIS now at the Keck~I telescope. The finest of the spaxel scales of
OSIRIS, with lenslets subtending 20~mas was used under conditions of exceptionally
good seeing, with the Canada-France-Hawaii Telescope (CFHT) seeing monitor reporting average seeing at 
optical wavelengths of $\approx$ 0$\farcs$3.
A third set of data was obtained on 2013 November 22 and 23 (MJD 56618.5 and 56619.5, midpoint=2013.893) 
with OSIRIS at the Keck~I telescope.
The seeing was not quite as good as in the year before, but still, very good data were obtained that
are well suited for an astrometric comparison with the 2012 data.


\begin{table}[htbp]
\begin{centering}
\begin{tabular}{|ccccccc|} \hline 
{\em Date} & {\em Telescope} & {\em Grating} & {\em Filter ($\lambda$ range)} & {\em Scale} & {\em Int. Time} & {\em FRP}\\ \hline
2011 Aug 21 & Keck II & old & Kn2 (2.04 - 2.14) & 50mas & 1$\times$300s + 1 sky & old\\
2011 Aug 21 & Keck II & old & Kbb (1.97 - 2.38) & 100mas & 1$\times$300s + 1 sky & old\\
\hline
2012 Nov 04 & Keck I & old & Kn2 (2.04 - 2.14) & 20mas & 4$\times$600s + 1 sky & new\\
2012 Nov 04 & Keck I & old & Hn3 (1.59 - 1.68) & 20mas & 8$\times$600s + 2 sky & new\\
\hline
2013 Nov 22 & Keck I & new & Kn2 (2.04 - 2.14) & 100mas & 4$\times$300s + 1 sky & new\\
2013 Nov 22 & Keck I & new & Kn2 (2.04 - 2.14) & 20mas & 4$\times$600s  + 1 sky & new\\
2013 Nov 22 & Keck I & new & Hn3 (1.59 - 1.68) & 20mas & 4$\times$600s  + 1 sky & new\\
\hline
2013 Nov 23 & Keck I & new & Kn2 (2.04 - 2.14) & 20mas & 4$\times$600s  + 1 sky & new\\
\hline
\end{tabular}
\caption{Log of the Keck OSIRIS Observations}
\label{sim-params}
\end{centering}
\end{table}

On 2012 Nov. 04 and 2013 Nov. 22, we obtained data sets with 600 s individual exposure time in the Hn3 filter,
covering the 1.644002~$\mu$m a$^{4}D_{7/2}$ -- a$^{4}F_{9/2}$ emission of [FeII] (referred in the
following as the 1.644~$\mu$m line) and also the fainter [FeII] $a^{4}D_{3/3}$ -- $a^{4}F_{7/2}$ line at
1.599915~$\mu$m (in the following called the 1.600~$\mu$m line). All wavelengths given in this paper 
refer to vacuum
and the line wavelengths and identifications are based on \citet{Aldenius2007}. 
Fig.~2 shows a raw spectrum of SVS~13 in the Hn3 filter, 
averaged over a large 0$\farcs$2$\times$0$\farcs$2 box centered on SVS~13, to include
the radiation from the object, but also the OH airglow sky flux used for wavelength calibration.
The [FeII] 1.600~$\mu$m line was evaluated and
generally corroborates the conclusions drawn from the 1.644~$\mu$m line, but due to its lower signal-to-noise ratio, these 
data are not presented here nor did we try to extract information on excitation conditions from
a comparison of those two [FeII] lines. 
\citet{Takami2006} have already presented a detailed study of the excitation conditions of H$_2$ and
[FeII] emission in SVS~13.
In addition to these shock-excited forbidden lines, 
OSIRIS in 4$^{th}$ grating order with the Hn3 filter also covers two atomic hydrogen emission lines, 
the 13-4 (1.611373~$\mu$m) and 12-4 (1.641170~$\mu$m)
recombination lines of the Brackett series \citep{Kramida2012}.
In the Kn2 filter, 4 data sets with 600~s exposure time each were obtained that 
cover the H$_2$ 1--0 S(1) line at 2.121833~$\mu$m \citep{Bragg1982}.

\begin{figure}
\begin{center}
\figurenum{2}
\includegraphics[scale=0.9,angle=0]{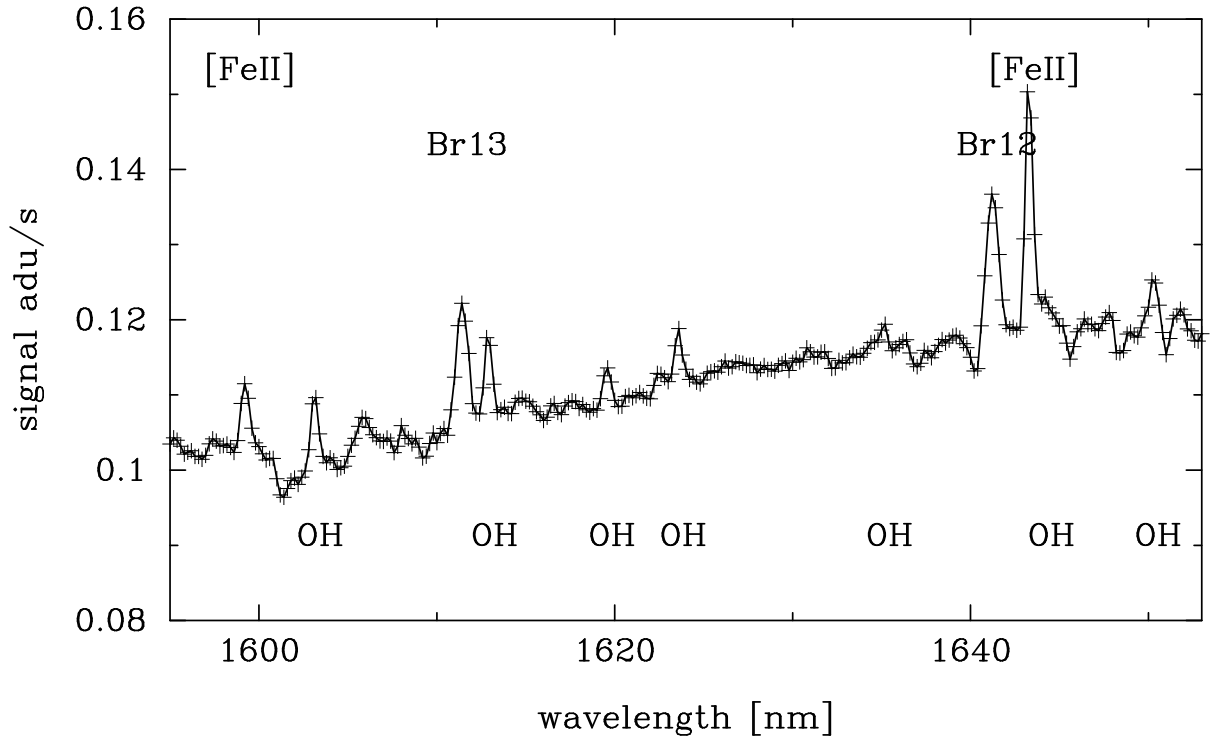}
\caption{
Raw spectrum extracted from the OSIRIS data cube in a 0$\farcs$2$\times$0$\farcs$2 box containing SVS~13, 
but also substantial sky OH airglow flux.
This figure serves to illustrate the Hydrogen Br~12 and 13 lines, the [FeII] lines, 
and the night-sky OH airglow lines
used for wavelength calibration.
}
\end{center}
\end{figure}

\subsection{IRIS Photometric Monitoring}
For the monitoring of SVS~13 at infrared wavelengths, the
brightness of the object around K=8 is actually a problem. Most archival infrared
images obtained with large telescopes and modern infrared cameras are saturated on SVS~13.
From 2012 to 2014, we have therefore monitored the NGC~1333 region, including SVS~13, with the
Infrared Imaging Survey (IRIS) system specifically to study the variability of
that object at the present epoch. IRIS is a 0.8~m telescope and 1024$\times$1024 infrared
camera dedicated to the monitoring of infrared variability. It has been described in
detail by \citet{Hodapp2010}. 

The IRIS infrared camera, which is a refurbished version of the UH Quick Infrared Camera (QUIRC) described
by \citet{Hodapp1996QUIRC}, is operated in a mode where after the first non-destructive read of the detector
array we immediately take a second read, and compute a double-correlated
image with only 2~s exposure time by differencing these first two reads. 
After this second read-out, the integration continues up
to a third read of the detector array, which gives the full integration time of 20~s
usually used for our monitoring projects. The first (2~s effective integration time)
images are far from saturating on SVS~13 and were used for the photometry presented here.
The IRIS camera raw data are processed in a reduction pipeline based on 
the Image Reduction and Analysis Facility (IRAF) software \citep{Tody1986}. 
The astrometric solution of the co-added images is calculated based on SExtractor \citep{Bertin1996} and
SCamp \citep{Bertin2005}. Photometry is obtained with the IRAF task PHOT and calibrated against the 2MASS
catalog \citep{Skrutskie2006}.

\section{Data Reduction}
\subsection{Keck OSIRIS Data Reduction Pipeline}
We chose to correct dark current and other
signal-independent detector artifacts in OSIRIS
by subtracting the median of 10 dark frames of 600~s exposure time
from each object exposure, 
in order not to waste observing time on frequent observations of an empty sky field
for sky background subtraction. 
The 2D detector frames were then processed into 3D spectral
data cubes with the OSIRIS final data reduction pipeline (FRP). 
This process of extracting the spectra from the raw detector format
is obviously critical for the wavelength calibration of the spectra
and therefore for the extraction of velocity information.
In 2013, the Keck Observatory released a new version of the data reduction pipeline 
(DRP v3.2) 
with newly calibrated ``rectification matrices'' that contain the parameters for the
extraction of individual spaxel spectra from the raw detector format into 
wavelength calibrated spectra. A first inspection of the newly reduced 2013 data
showed better consistency of the wavelength calibration across all spaxel and
therefore the 2012 data were also re-reduced with the new software and calibration data.
These newly reduced data did indeed change the velocity structure observed in the 
SVS~13 jet to the point of forcing a different conclusions.

The individual data cubes in the Hn3 filter were cleaned of spurious bright spaxels using the
IRAF COSMICRAY task on individual planes of the data cube. 
All data cubes in the Hn3 filter were spatially mosaicked together with offsets determined
by the centroid of the bright continuum image of the SVS 13 star.
For the shorter wavelengths, this method resulted in superior image quality
compared to using offset data supplied by the AO system.
For the longer wavelength spectra in the Kn2 band, where we only took one dither
pattern, relying on the offsets
provided by the laser guide star system proved sufficient.
For both the Hn3 and Kn2 data cubes, the individual planes (wavelengths) of the 
resulting combined spectral data cubes
were adjusted for zero sky background using
the background zero adjustment in the IRAF task IMCOMBINE. The primary purpose
of this procedure was to remove the OH(5,3)R$_{1}$(2) night sky line at 1.644216~$\mu$m
that partly overlaps with the [FeII] 1.644002~$\mu$m line, as was discussed by \citet{Davis2003}.
The integral field spectrograph OSIRIS records all wavelength planes simultaneously,
so that the PSF at continuum wavelengths is recorded under identical atmospheric
conditions than the structure at emission line wavelengths. Over small wavelength intervals
where the wavelength-dependence of diffraction is negligible, very precise continuum
subtraction can be achieved.
The continuum adjacent to a given line was computed as the average of approximately
10 wavelength planes outside of the line profile and centered on the line.

\subsection{OSIRIS Spaxel Scales and Image Orientation}
Between the 2011 and 2012 runs, the OSIRIS instrument was moved from the Keck~II telescope to
Keck~I. Both telescopes and their adaptive optics systems have nominally identical focal lengths,
but for an astrometric comparison of data from these two telescopes, the spaxel scale in OSIRIS
should be observationally verified.
The best data available for this purpose were kindly provided by T. Do and J. Lu
from their work on astrometry of stars near the Galactic Center where they calibrated OSIRIS astrometry relative
to astrometry with NIRC2, an instrument that stayed at the Keck~II telescope. 
These data indicate that at Keck I, the spaxel scale in mas~pixel$^{-1}$ 
of OSIRIS is 1.28\% larger than it had been at Keck~II. 
While installed on Keck~I, between our 2012 and 2013 observing runs, the grating of OSIRS was also
changed, resulting in a new wavelength calibration. 
Also, as was discussed above, the data reduction pipeline used at
Keck for OSIRIS data was upgraded.
Finally, in moving from Keck~II to Keck~I,
the optical path leading to OSIRIS now contains one less mirror reflection, leading to a change in
parity of the images. We used the OSIRIS camera with its wider field to obtain a few images tying
the OSIRIS spectral data cubes astrometrically to conventional images of SVS~13 and one detectable
star in its vicinity. In addition, we observed other objects to ascertain that our data are presented
in the correct parity and orientation on the sky.

The precise scale ratio of the nominal 50~mas (used in 2011) and 20~mas (used in 2012 and 2013) spaxel scales of OSIRIS was measured on
setup data cubes obtained in 2008 while observing the $\approx$ 0$\farcs$25 separation 
binary $\sigma$ Orionis \citep{Hodapp2009}.
By averaging the component separation measured in each continuum
data cube plane within $\pm$~0.1~$\mu$m of 2.122~$\mu$m, a ratio of 2.4534~$\pm$~0.0005 between the two spaxel scales
was measured. This value deviates by about 1~\% from the nominal scale ratio of 2.5. It should be noted that
these deviations from the nominal spaxel scale ratios and the, even smaller, errors of this measurement are much smaller
than the astrometric effects of shock-front motion and expansion discussed later in this paper.

\subsection{Lucy-Richardson Deconvolution}
The continuum images near the emission line features show nothing but the
unresolved stellar object SVS~13. They are thus suitable as a PSF kernel for
deconvolution of the individual (wavelength) cube planes. The deconvolution
was done with the Lucy-Richardson algorithm \citep{Richardson1972} and \citep{Lucy1974}, 
as implemented in the IRAF STSDAS package. How fast the Lucy-Richardson algorithm
converges depends on the signal-to-noise ratio in the frame. We have tried out
different numbers of iterations and chi-square criteria and have chosen a combination
that leads to a significant improvement in the spatial resolution without creating
obvious artifacts. 
Working with
the original data without continuum subtraction led to strong artifacts in the wings of the
overwhelmingly bright stellar source. 
For the deconvolution of faint emission features well separated
from SVS~13 we have therefore worked from the continuum-subtracted data. 
For the purpose of the grey-scale and RGB representations of the images, we have
computed a ``high-dynamic-range'' version of the image by adding the deconvolved
continuum-subtracted image and a small fraction of the image flux of the original
deconvolved image, effectively creating an image where the continuum source is 
strongly suppressed, but still indicated in the image.
Fig.~3 shows the S(1) image from 2011 in greyscale, and false color representations 
of the deconvolved ``high-dynamic-range'' 2012 and 2013
data in the H$_2$ S(1) line (red), [FeII] 1.644$\mu$m (green) and continuum (blue).
Fiducial marks are included to illustrate the expansion of the S(1) bubble over
the course of those 2 years.

\begin{figure}
\begin{center}
\figurenum{3}
\includegraphics[scale=1.00,angle=0]{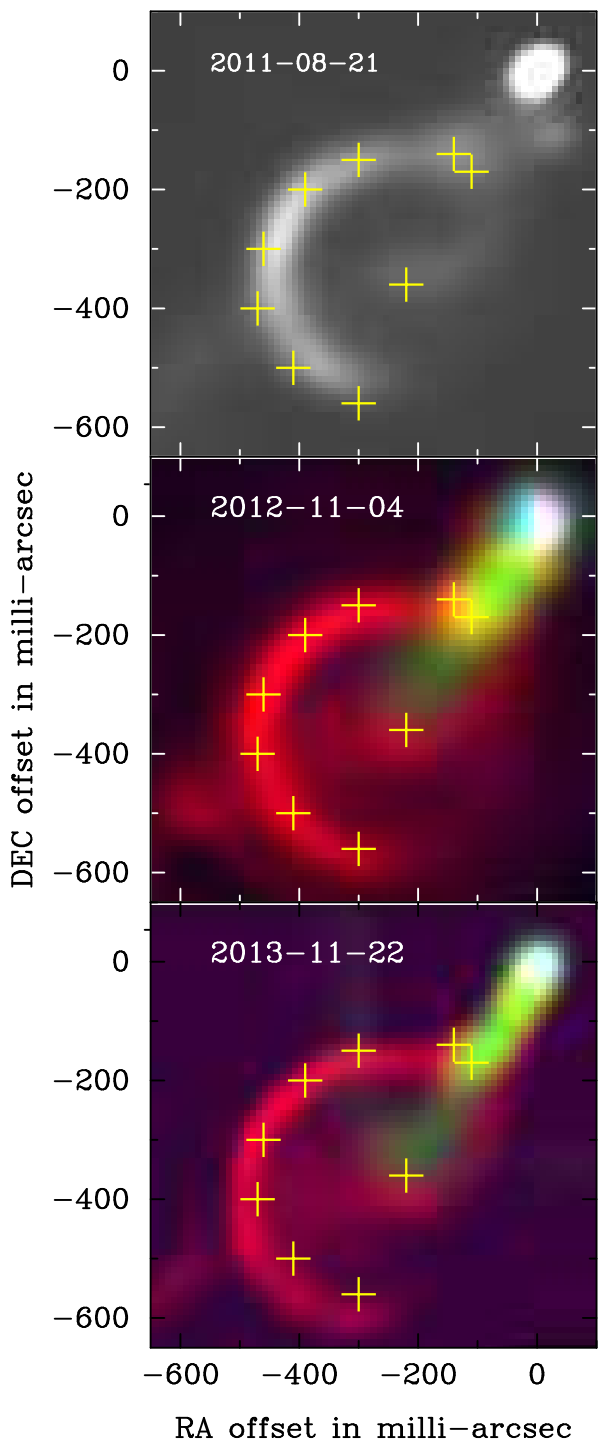}
\caption{
The SVS~13 bubble in 2011, 2012, and 2013. 
False color images composed of the deconvolved and continuum-subtracted H$_2$ S(1) image
(red channel) and the continuum subtracted [FeII] emission line image from 2012 Nov. 4. All three images are
aligned to better than 1 mas, even though the deconvolution residuals in the continuum subtracted images do
not align precisely. 
To mark the position 
of the continuum source SVS~13,
the blue channel shows 10\% of the 
continuum near the [FeII] line.
A few fiducial marks indicate positions
of the bubble rim in 2012. The comparison of these marks to the images taken earlier
and later clearly indicated the expansion motion of the bubble.
}
\end{center}
\end{figure}

\subsection{Spectral Resolution and Wavelength Recalibration}
The OSIRIS final data reduction pipeline (FRP) extracts the data from the raw 2-dimensional 
detector frame and forms a spectral data cube with 0.2~nm spacing in the H band (4$^{th}$ grating order) and
0.25~nm in the K band (3$^{rd}$ grating order).
The OH airglow lines in the bandpass of the Hn3 
and Kn2 
filters give an opportunity to check the wavelength calibration
and to measure the spectral resolution
In particular, the 
OH(5,3)R$_{1}$(2) line at 1.644216~$\mu$m is very close to the wavelength of the [FeII] 1.644~$\mu$m line.
For the analysis of the [FeII] line emission, 
the uniformly distributed OH emission was eliminated by adjusting the spatial
median of the flux in the image to zero. In contrast, for the wavelength calibration,
the original data were used and the flux in the OH line was measured. To account
for continuum stray light, a continuum subtraction using the same continuum as
for the [FeII] line extraction was used. 
The individual OH airglow lines
were measured by summing up the flux of 400~pixels without detectable object flux,
and measuring the position and width of the OH airglow lines in the resulting
one-dimensional spectrum using the IRAF task SPLOT. We used the vacuum wavelengths of
OH lines given by \citet{Rousselot2000} and found that the wavelength calibration
provided by the OSIRIS reduction pipeline required a slight correction in zero point
and dispersion, of order 0.01~nm. After this correction, the individual OH line positions have residual
wavelength errors of 0.008~nm, about a factor of two larger than the fit residuals
given by our wavelength reference, \citet{Rousselot2000}. 
The OH airglow lines are, in reality, close doublets with line spacing
well below the resolution of OSIRIS. 
The measured FWHM of the OH lines was 0.44~$\pm$~0.06~nm in the Hn3 filter,
and
0.61~$\pm$~0.06~nm in the Kn2 filter,
a little more than two wavelength planes in the OSIRIS data cubes. 
Near the [FeII] line at 1.644~$\mu$m, one wavelength interval of the data cube
corresponds to 37~$kms^{-1}$ in radial velocity. The spectral resolution near that
line therefore corresponds to 81~$kms^{-1}$.
Near the H$_2$ 1--0 S(1) line at 2.122~$\mu$m, one wavelength interval of 0.25~nm
corresponds to a radial velocity difference of 35~$kms^{-1}$, and the measured 
FWHM of the spectral lines corresponds to 85~$kms^{-1}$ width of the spectral
profile. Most of the velocity effects discussed in this paper are therefore smaller than
one spectral resolution element.
The accuracy of velocity measurements depends on the signal-to-noise ratio
of the emission in question. In our discussion of the velocity structure of
the [FeII] microjet in section 4.5, we will present a detailed analysis of
the measurement errors in that particular case.

\subsection{No Telluric Absorption Correction}
We have taken data on an A0V star during the observations, but in the end have
chosen not to use these for telluric absorption correction. Our data analysis
does not rely on spectrophotometric correction of the data cubes. In the Hn3 filter,  
the study of the jet in the 1.644~$\mu$m [FeII] line
and of the HI13-4 and HI12-4 Brackett lines are potentially all affected by the broad atomic hydrogen absorption
in the standard star, and the data reduction pipeline would therefore interpolate through the
hydrogen lines, making the telluric absorption correction meaningless at these wavelengths.

We have also not been able to use those standard star measurements for a flux calibration
of our spectra. A consistency check of data obtained with different exposure times showed
a problem with the way OSIRIS computes the effective integration times for multi-sampled
exposures. We do not have the data to fully calibrate this effect, and given that the 
flux calibration is not essential for our data analysis, prefer not to give a flux
calibration for our data.

\section{Results and Discussion}
\subsection{The SVS~13 Outflow on Different Spatial Scales}

Figure 1 shows the H$_2$ shock-excited line emission associated with SVS~13 on four different
spatial scales: The top image is an archival Spitzer Space Telescope Infrared Array Camera (IRAC) \citep{Fazio2004}
channel 2 (4.5~$\mu$m) 
image covering most of the outflow emission. The second panel is a small cutout from a ground-based,
seeing limited image in the H$_2$ 1--0 S(1) line obtained at the UH 2.2m telescope,
showing just the emission in the immediate vicinity of SVS~13. 
The third panel from the top is
the S(1) integrated line image obtained with the 100~mas scale of OSIRIS in 2011 and the bottom
panel shows the 20~mas
scale S(1) image from OSIRIS obtained in 2012, without deconvolution. 
We show this figure to demonstrate the relationship between the bubbles in
the SVS~13 outflow very close to the source and the larger scale structure of the flow further
downwind. This relationship is not entirely trivial, since not all the bubbles have propagated
in the same direction from the SVS~13 star.
The seeing-limited ground-based image shows the brightest of these bubbles in the glare of the
PSF. The intermediate, 100~mas scale, OSIRIS line image shows three distinct partial bubbles
near the source of the outflow, the outer two of which were also observed, but not further
discussed, by \citet{Noriega2002}, on HST/NICMOS images and had also been detected through
UKIRT narrow-band Fabry-Perot imaging by \citet{Davis2002}. The best prior detection of these
bubbles was by \citet{Davis2006} who used adaptive-optics-assisted long-slit spectroscopy and
detected all the bubbles as a series of emission maxima along the slit.

The most recent emission, the 0$\farcs$2 long [FeII] jet, its faint extension into the S(1) bubble, 
and that youngest S(1) bubble are oriented at P.A. 145$^{\circ}$, as seen in Fig. 3.
The older two bubbles, visible in the 100~mas image of Fig.~1, are lying
more to the south at P.A. 159$^{\circ}$ than those aforementioned features, indicating some variations in the
jet emission direction.
In contrast the main chain of Herbig-Haro knots HH~7-11 lies at P.A. 123$^{\circ}$ \citep{Davis2001}. 

While at optical wavelengths, the SVS~13 outflow manifests itself only in the
Herbig-Haro chain HH~7-11 \citep{Herbig1983}, longer wavelengths increasingly reveal a counterflow that appears displaced
from the HH~7-11 axis, and appears in general fainter and less organized (Fig.~1, top panel). Based on Spitzer
telescope images over a time span of 7~years, \citet{Raga2013} have obtained one proper
motion data point for a relatively well defined knot in that counterflow, and this one point is consistent
with the counterflow originating from SVS~13. Also, no other embedded protostar has been detected
by \citet{Walsh2007}
to the north of SVS~13 that might explain this flow as being independent from SVS~13.
The displacement in the outflow axis, in combination
with the recent changes in outflow direction reported in this paper, suggest that the source
of the SVS~13 outflow is changing direction, probably due to some precessing motion,
as will be discussed in more detail in 
section 4.2.
As an explanation for the differences between the HH~7-11 flow and the counterflow, \citet{Walsh2007}
have suggested that the northern counterflow enters into the central cavity of NGC~1333, resulting in
different ambient pressure and excitation conditions than the HH~7-11 flow.

\subsection{H$_2$ Shock Front Proper Motions}
We have obtained astrometry of the bubble expansion on two different spatial scales.
Relatively wide field OSIRIS data cubes with the 100~mas spaxel$^{-1}$ scale were obtained
on Keck~II on 2011 August 21 and Keck~I on 2013 November 23. A difference image of the H$_2$~S(1) emission
at these two epochs is shown in Fig.~4 and illustrates that all three of the closest shock fronts
to SVS~13 are showing noticeable motion. The farthest of these three shock fronts was defined enough
to allow a cross-correlation measurement of its expansion age in the box indicated in the figure.

\begin{figure}
\begin{center}
\figurenum{4}
\includegraphics[scale=0.9,angle=0]{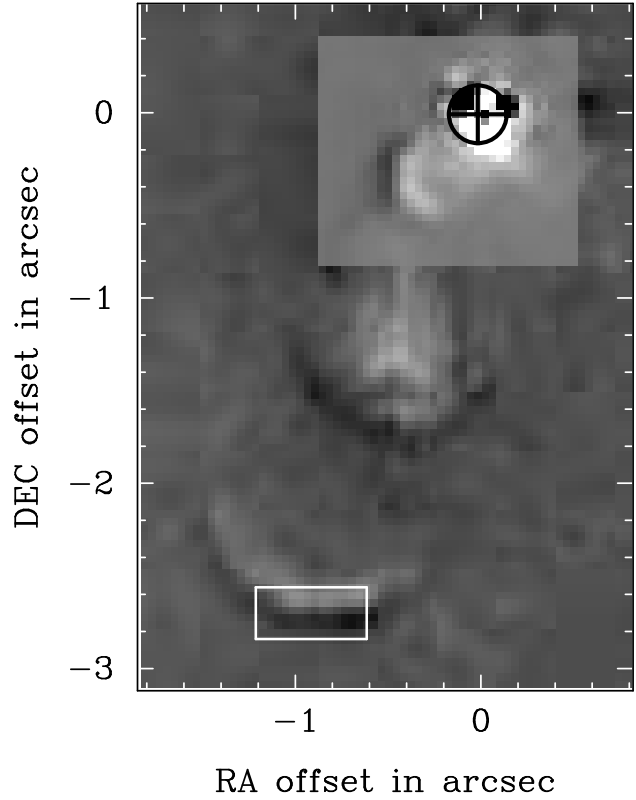}
\caption{
Difference frame of the 100~mas scale images taken in 2011 and 2013. The cross-correlation box for the most
distant shock front is indicated.
The insert shows the inner (youngest) bubble and the subtraction residuals of SVS~13 in different flux scaling.
The scaling is such that bright pixels in the frame show more flux in 2011 than in 2013. All three shock
fronts therefore show the signature of expansion with the dark (2013) features farther away from SVS~13 than
the bright (2011) features.
}
\end{center}
\end{figure}

Astrometry of the smallest and youngest expanding bubble was done on the 2012 November 4 and 2013 November 22/23 data 
that were taken at the Keck~I telescope with the same adaptive optics system and the
same spaxel scale (20~mas spaxel$^{-1}$) and are therefore suitable for a precise astrometric measurement.
In this case, the individual planes of the data cube were deconvolved using the Lucy-Richardson
algorithm, to improve the definition of the bubble edge.
Figure~3 illustrates the expansion and motion of the youngest, smallest bubble between the 
three epochs by superposing fiducial marks that outline features in the middle (2012) image.

For fine registration of the images, the IRAF task XREGISTER was used to measure the relative
alignment of all the frames at the position of the SVS~13 stellar object and a magnified
and optimally registered version of the images was produced. To measure the expansion of the
bubble, we worked under the assumption that its motion relative to the star can be described
as a simple constant velocity expansion with the origin at the position of the star. We computed magnified versions
of the individual 2012 November 4 images with magnification factors in the range from 0.95 to 1.10 and the center
of the magnification on the SVS~13 star. 
We then computed the product 
of these magnified images and each of the individual 2013 November 22 images.
The average of this product in a box centered on the S(1) bubble feature image varies smoothly with expansion factor
for each of these pairs and the maximum of this one-dimensional correlation function was simply read from
the table of cross-correlation values. We show the cross-correlation functions of the 2012 and 2013 high resolution
images in Fig.~5 to illustrate the
variations due to noise and deconvolution noise amplification, and to document how the errors of
this measurement were obtained.

\begin{figure}
\begin{center}
\figurenum{5}
\includegraphics[scale=0.9,angle=0]{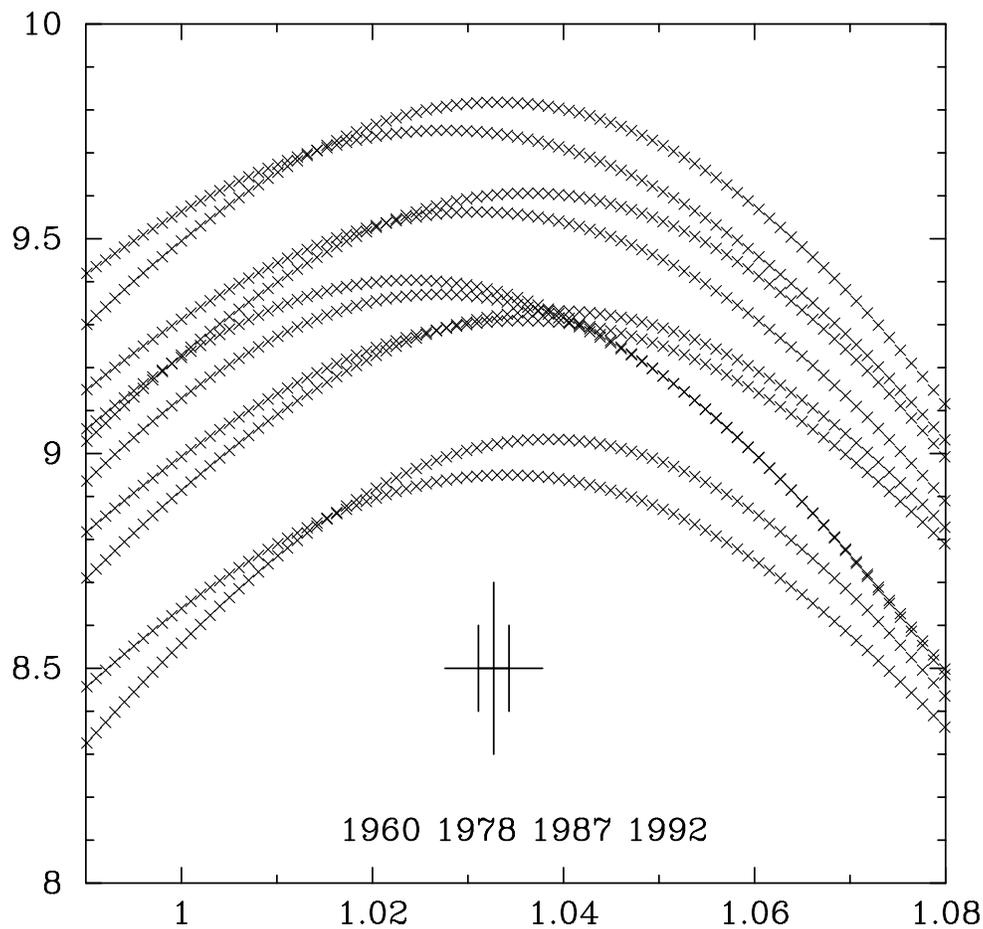}
\caption{
The 10 individual cross-correlation results between individual 2012 and 2013 images taken with
the OSIRIS 20~mas spaxel scale. The horizontal axis is the expansion factor of the 2012 data to match
the 2013 data, and the vertical axis is the average of the product of the two frames, i.e., the 
cross-correlation function. The expansion starting dates corresponding to the expansion factors are also 
indicated. The error bars indicate the 1~$\sigma$ error of the mean scaling factor.
The mean value corresponds to a kinematic start of the expansion in 1980, an early limit to the true
starting date since the expanding bubble will realistically be decelerated.
}
\end{center}
\end{figure}

The expansion factor so determined was then converted to a kinematic expansion age of 32 yrs (prior to 2012), i.e.,
kinematic starting time of the expansion in 1980, which
gives the upper limit to the true age of the bubble assuming that the bubble has been expanding at 
constant velocity. Since, realistically, the bubble is expanding into the dense environment of a molecular core,
the true age of this youngest bubble will be smaller than the kinematic expansion age and is therefore consistent with this
expanding bubble having been generated in the 1990 photometric outburst of SVS~13 that will be discussed
in secion 4.4.
In 2012, the apex of the bubble was located 654~mas from the star. The front end of the bubble
therefore has moved with an average projected angular velocity 
of 20~mas~yr$^{-1}$ (4.7~AU~yr$^{-1}$ = 22.3~km~s$^{-1}$ at 235 pc distance) away from the
star. Similarly, in 2012, the bubble had a radius of 212 mas and an average radial proper motion
of 6.6~mas~yr$^{-1}$ (1.55~AU~yr$^{-1}$ = 7.35~km~s$^{-1}$).
The proper motion of the bubble center is therefore the proper motion of the
leading shock front minus the radial expansion: 13.4~mas~yr$^{-1}$ $\approx$ 15~km~s$^{-1}$.

In order to also obtain an estimate of the kinematic age of the next two more
distant bubbles, we used data cubes obtained in 2011 August 21 on Keck~II with
the 100 mas scale, and similar measurements obtained on 2013 November 22 on Keck~I.
The difference of these two images is shown in Fig.~4.
The only shock front where a clear maximum of this correlation function was detected
in these 100 mas data was the most distant of the three, indicated by a box in Fig.~4, separated from the star
by 2.87$^{\prime\prime}$ in 2011. For this bow shock front,
we derive a kinematic formation time of 1919~$\pm$~7 and a linear projected proper 
motion of 31~mas~yr$^{-1}$.
We have not obtained a reliable kinematic age for the middle shock-excited 
feature at $\approx$~1$\farcs$5 from the star, between
those two features discussed above (Fig. 4). The fact that this rather poorly defined system
of shock fronts lies pretty precisely in the middle of the 1980 and 1919 (kinematic 
formation time) features suggests that this feature must have formed, again in the
kinematic sense without accounting for deceleration, around 1950.
If there is a regular pattern to the formation of these bubbles, which with only three
examples cannot be convincingly established yet, the next outburst could be expected
within the next decade.

It should be noted, as was already pointed out by \citet{Khanzadyan2003}
that the proper motion of the major shock fronts in the older parts of the SVS~13 outflow indicate a much
longer time interval between shock front generating events: about 500~yrs.
It is not clear whether SVS~13 exhibits multiple periods, or whether the frequency
of ejection events has recently increased. An argument for the latter point of 
view may be that the outflow direction has apparently changed in the past
century, as we will discuss now.

As demonstrated by Fig. 1, the two previous bubbles are located more toward the SE of
SVS~13 and, for example, \citet{Davis2006} chose P.A. 159$^{\circ}$ as the best slit orientation
to cover them. The larger chain of HH objects 7 to 11 is oriented along a position angle of 123$^{\circ}$
\citep{Davis2001}.
The larger scale proper motion study by \citet{Raga2013} based on Spitzer 4.5 $\mu$m images shows some emission knots to
the SE of SVS~13 with a proper motion vector generally to the SE, in particular the Herbig-Haro
knots 7, 8, and 10. North of SVS~13, in the counter-flow, they find an emission knot with a generally northern 
proper motion vector (P.A. -10$^{\circ}$) that they ascribe to a chance superposition of another
outflow far to the south of SVS~13 and with generally northern outflow direction.
The [FeII] jet originating from SVS~13 (Fig.~3) is oriented along P.A. 145$^{\circ}$ and the H$_2$ bubble
center is displaced from SVS~13 along the same angle.
With the exception of this most recent H$_2$ bubble,
the other recent mass ejection events from SVS~13 have therefore ejected material initially
in a more 
southerly 
direction (P.A. $\approx$ 155$^{\circ}$ - 159$^{\circ}$), as seen in Figs. 1, and 4. 
With outflows generally being bipolar, there must also be mass 
ejected into a northerly direction. We therefore believe that, contrary to the assertion
by \citet{Raga2013}, the 4.5~$\mu$m emission knot
north of SVS~13 and with northerly proper motion is part of the counter jet to the
bubbles reported here. The more distant parts of the S(1) emission NW of SVS~13 are anti-parallel,
but laterally displaced, from the HH~7-11 system of emission knots. We suggest that
the HH~7-11 chain, the system of bubbles immediately south of SVS~13, the emission knots north
of SVS~13, and the more distant emission knots further to the NW (Fig. 1) are all part of the same
bipolar outflow originating in SVS~13. This outflow has the S-shaped morphology indicative of a 
precessing or otherwise unstable jet source. 
Corroborating this, radio interferometry mapping of molecular emission near SVS~13 by \citet{Bachiller2000} has
similarly found an orientation of high-velocity material south of SVS~13 different than the HH~7-11
Herbig-Haro chain. They had already concluded that the differences in the alignment 
of features of different age indicate a precessing source of the outflow.
A prominent other example of such S-shaped morphology of a molecular hydrogen jet, IRAS~03256+3055,
is located just south of SVS~13 in NGC~1333 and has been studied in detail by
\citet{Hodapp2005}.

The relatively low proper motion and spatial velocity measured for the H$_2$ bubble studied
here is consistent with proper motion measurements of the more distant HH~7-11 chain of
shock fronts by \citet{Herbig1983}, \citet{Noriega2001}, and \citet{Raga2013}, but is 
inconsistent with the much higher proper motions reported by \citet{Chrysostomou2000} in
the first near-infrared proper motion study of HH~7-11.

While a precessing accretion disk provides an explanation for the rapid changes in the
outflow direction observed in SVS~13, our data do not show any indication for the presence
of a companion object that would be close enough to cause the disk precession
on the timescales discussed here. 
From the size and location of the H$_2$ bubbles, we can conclude that bubble ejection events happen with a period of
several decades. In the model where such events are triggered by periastron passages of a 
companion object, the orbital semimajor axis must be of order of tens of AU, or several of the
original 20~mas spaxels of our data. The fact that we don't see a companion object implies
that such an object, if it existed, is intrinsically too faint and/or too deeply embedded to
be visible in the H and K atmospheric windows.

The only other case of a young stellar object with a pronounced bubble structure of
its outflow is XZ~Tau that was studied in detail by \citet{Krist2008} on the basis of
optical multi-epoch HST imaging. Very similar to the case of SVS~13 discussed here,
their images show a series of bubbles with a collimated jet propagating inside of the bubbles.
\citet{Krist2008} reported initial results of numerical simulations of a very
young pulsed jet in close proximity to its driving source. Their simulations were
specifically tuned to reproduce the observations of the XZ~Tau~A chain of bubbles
and therefore modeled a faster, more rapidly pulsing jet resulting in more rapidly
expanding, overlapping bubbles.
The general scenario underlying their model is however applicable to our case of SVS~13:
The FUor-like photometric outburst in 1990 created a short-lived pulse of jet activity. This newly created,
relatively fast jet ran into slower moving material ejected prior to the outburst event,
and this internal shock created an expanding ``fireball'' that subsequently expanded
ballistically into a bubble carried away from the star by the outflow.
Some time after the formation of the bubble, a fast continuous jet then emerges
to catch up with the bubble, pierce it, and partially destroy it.
In SVS~13 this process currently repeats itself about every 30 years, creating the series
of bubble fragments that forms the string of Herbig-Haro objects.
In distinction from XL~Tau, the case of SVS~13 also involves a significant change in
the direction of the jet and bubble ejection leading to the S-shaped overall morphology
of the Herbig-Haro chain.

\subsection{H$_2$ Shock Front Radial Velocity}
The individual velocity channels of the H$_2$ S(1) line emission
are shown in Fig.~6. In the blue-shifted wing of the velocity distribution, emission projected
on the center of the bubble is visible, which is the expected characteristic of an expanding 3-dimensional
bubble rather than a 2-dimensional ring.  In Fig.~7, a color-coded velocity map of the H$_2$~1--0 S(1) line emission is presented.
The S(1) emission shows three distinguishable velocity features. Emission near the intersection
with the jet (traced in [FeII]) shows the smallest blueshifted velocities and is coded red.
The rim of the S(1) bubble where the line of sight is tangential to the bubble shows intermediate
velocities, coded yellow in the figure. 
The highest velocities towards the observer are measured in the filamentary features projected against
the center of the bubble and are coded in blue.
In the brightly visible rim, the velocity centroid varies between -40 and -55~km~s$^{-1}$
relative to the systemic velocity of the molecular material around SVS~13, 
while in the simple model of an expanding
shell, those velocities should be constant and representative of the center motion of the bubble. 
We take -47~$\pm$~7~km~s$^{-1}$
as the typical radial velocity of the bubble center. 

\begin{figure}
\begin{center}
\figurenum{6}
\includegraphics[scale=1.0,angle=0]{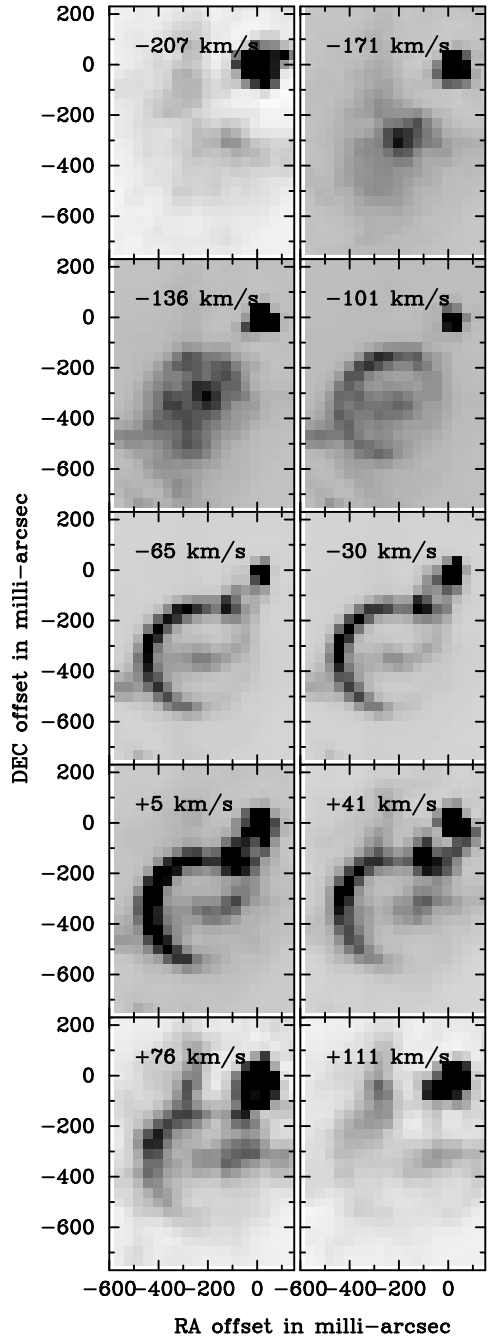}
\caption{
Continuum-subtracted H$_2$ 1--0 S(1) line images of the SVS~13 jet. The velocities indicated
in each panel are relative to the systemic velocity of the SVS~13 core. The bubble feature
is blueshifted relative to the systemic velocity.
}
\end{center}
\end{figure}

\begin{figure}
\begin{center}
\figurenum{7}
\includegraphics[scale=0.7,angle=0]{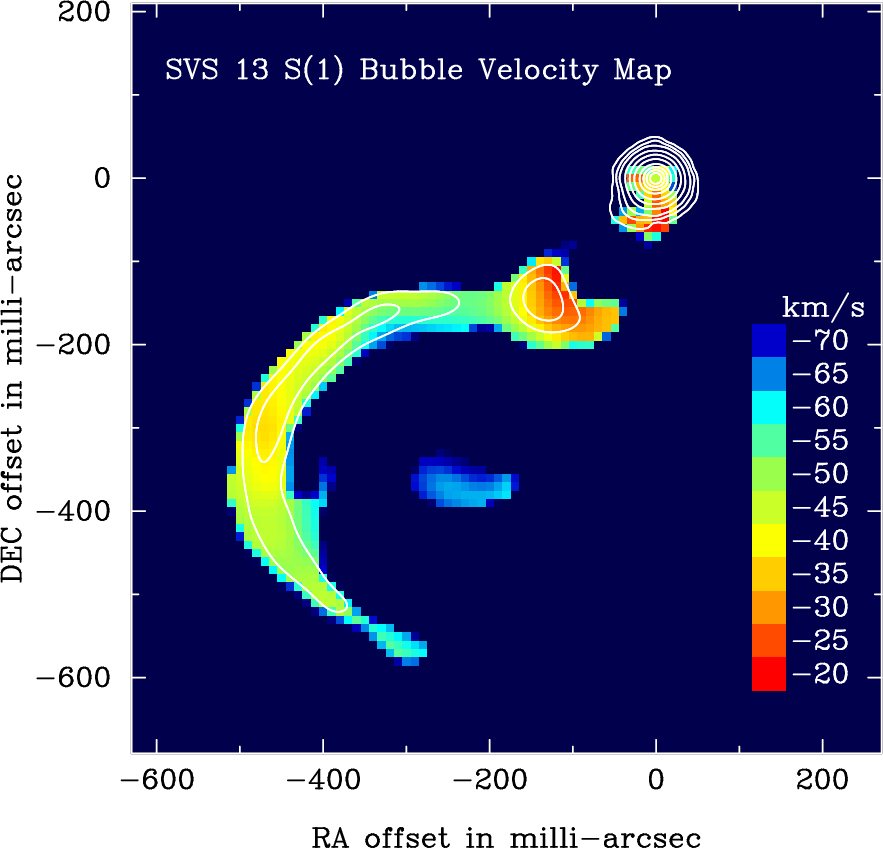}
\caption{
Velocity map of the SVS~13 bubble in the H$_2$ S(1) line at 2.122~$\mu$m. Three distinct velocity features can be
distinguished: The emission at the intersection of the jet with the bubble has the lowest (blueshifted) radial velocity,
The rim of the bubble has intermediate velocities, and the features projected against the center of the bubble
have the highest blueshifted velocities. Overlayed on the velocity map are contours of the high-dynamic range
Lucy-Richardson deconvolved
flux maps that indicate the position of SVS~13 itself.
A color version of this figure is available in the electronic version of this paper.
}
\end{center}
\end{figure}

With the proper motion of the bubble center of 15~$\pm$~2~km~s$^{-1}$, this
suggests an inclination angle of 18$^{\circ}$ $\pm$ 3$^{\circ}$ against the line of sight.
\citet{Takami2006} had given an inclination angle of 20$^{\circ}$ to 40$^{\circ}$
for emission close to SVS~13 while \citet{Davis2001} has given 40$^{\circ}$ for more distant
emission knots. 
All the measurements of the inclination angle were done on different shock features.
Their spread is therefore a combination of measurement uncertainties and the true
variations in the motion of these shock fronts.
Irrespective of which of the shock fronts are measures, all data indicate that
the outflow emerging from SVS~13 is pointed strongly towards the observer.
This explains why the counterjet, which moves away from the observer and into the molecular core
around SVS~13, is not detectable at optical wavelengths.
Shock fronts of this velocity running into stationary ambient
molecular hydrogen are certainly capable of exciting v=1--0 S(1) line emission and are not in danger of
dissociating the H$_2$, see, for example, \citet{Draine1980}.

\subsection{The Outburst and Light Curve of SVS~13}

The infrared source SVS~13 was discovered by \citet{Strom1976} at a K-band magnitude
of 9.08 in a 36$^{\prime\prime}$ aperture. Soon after the discovery, \citet{Cohen1980} reported
K~=~8.48 in a 30$^{\prime\prime}$ aperture observed in 1978. \citet{Liseau1992} reported an
observation by G. Olofsson from 1980 at K=8.7 in a 14$^{\prime\prime}$ aperture as a private
communication. \citet{Harvey1984} measured a pre-outburst brightness
on K=9.34 in a 6~-~8$^{\prime\prime}$ aperture at the IRTF on 1981, Oct. 11-13 and \citet{Cohen1983} measured
K=9.30 in a 16$^{\prime\prime}$ aperture on 1981, Dec. 3-9.

The object experienced a sudden increase in brightness
around 1990, as first reported by \citet{Mauron1991} and further studied by \citet{Eisloeffel1991},
\citet{Liseau1992}, and 
\citet{Aspin1994}. In the K band, where the best pre-outburst data 
are available, as listed above, the pre-outburst magnitude showed some variation
between 9.0 and 9.5~mag. Post-outburst, \citet{Aspin1994} documented brightness
variation between 8.0 and 8.6~mag. Based on the small amplitude of the brightness
increase, the post-outburst brightness fluctuation, and the emission lines in its
spectrum, both \citet{Eisloeffel1991} and \citet{Aspin1994} have concluded that
SVS~13 underwent an EXor or similar outburst, but both papers left the possibility open
that the outburst may be of a different nature.
\citet{Khanzadyan2003} has studied the photometric behavior of SVS~13 again and concluded
that the object had not returned to its pre-outburst brightness at that time. 
Motivated by the uncertain classification of this event,
we have re-examined the historical photometry and are discussing new measurements.
We have tried to gather the available photometric data on SVS~13 from the literature
and data archives. While many images of the SVS~13 region exist, the star SVS~13 is
saturated on most of these, and useful data can only be obtained from shallow surveys,
usually with small telescopes.
In Fig. 8, we show all the available photometric data as a light curve. 

\begin{figure}
\begin{center}
\figurenum{8}
\includegraphics[scale=0.9,angle=0]{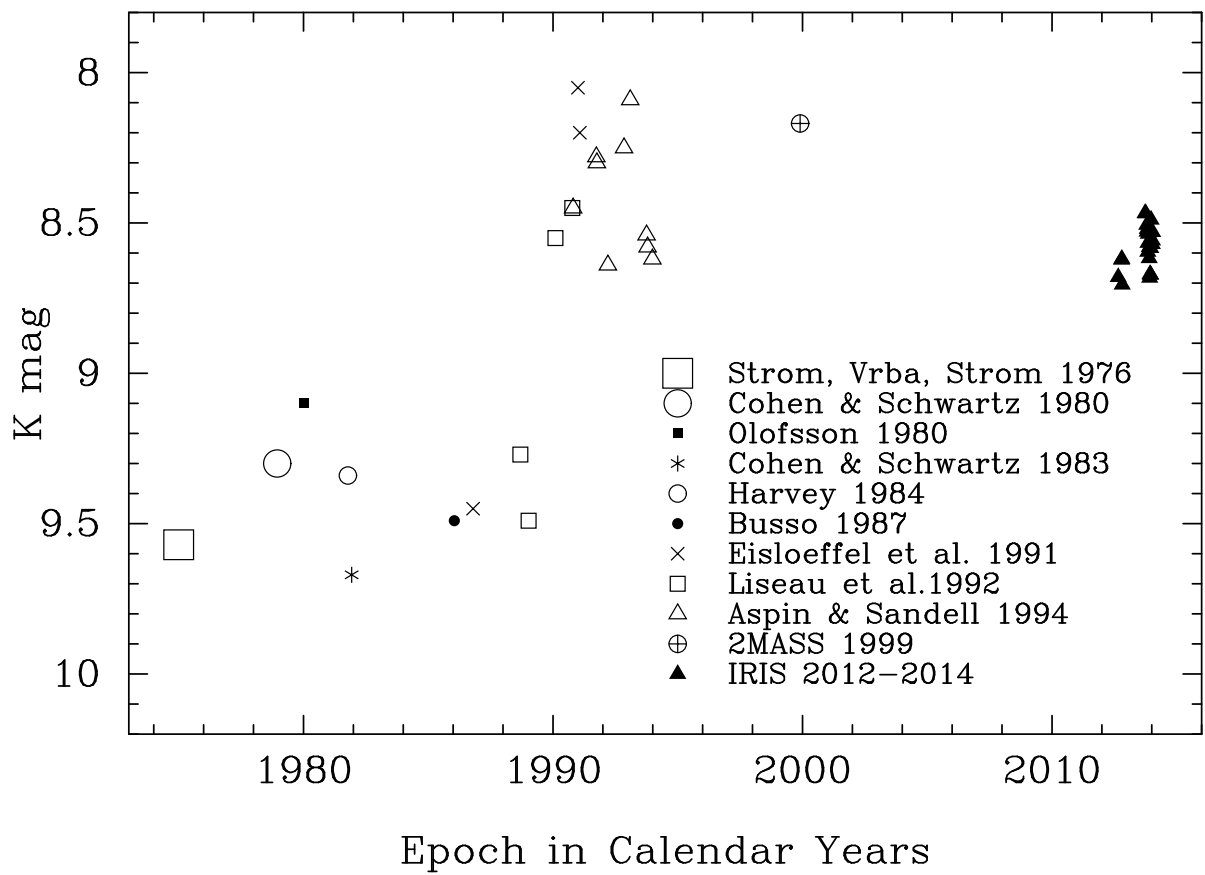}
\caption{
The lightcurve of SVS~13 from its discovery in 1974 to the present. This lightcurve is compiled from
all published K-band photometry of this source, the 2MASS catalog value, and recent photometry from
the IRIS telescope. The four oldest photometry data points have been corrected to match the much smaller
photometric apertures used in the more recent photometry based on the flux distribution in present epoch images.
The labels give the bibliographic reference, not the epoch of the observations.
}
\end{center}
\end{figure}

The 2MASS survey list SVS~13 as K$_s$=8.169 and these data had been obtained on
Nov. 26, 1999. Recent K$_s$ data from the IRIS telescope
\citep{Hodapp2010} show SVS~13
varying in the range of K$_s$=8.46 to 8.70
between 2012 August and 2014 January.
The photometric color transformations between the UKIRT system and the 2MASS system
for the K vs. K$_s$ filters are insignificant \citep{Carpenter2001}, so a direct comparison can be made
between the measurements by \citet{Aspin1994} and the most recent data. 
There is extended emission around SVS~13, so very large photometric apertures tend to
overestimate the brightness. 
The \citet{Strom1976}, \citet{Cohen1980}, \citet{Olofsson1980} and \citet{Cohen1983} 
data were corrected to the aperture diameter
of 6$^{\prime\prime}$ used for the IRIS photometry, while all other photometric
data shown here were originally obtained with apertures in the range of 5$\arcsec$ - 8$^{\prime\prime}$,
close enough to the IRIS aperture to not require a correction. 
Figure~8 shows these aperture-corrected photometric values and clearly demonstrates 
that SVS~13 has not declined back to its pre-outburst ($<$~1990) brightness,
but remains at or near its peak post-outburst brightness with some indication that brightness fluctuations
have diminished over the course of the past 24 years.

Similar to the case of the low-luminosity, deeply embedded, cometary nebula OO Ser
\citep{Hodapp1996} and \citep{Hodapp2012}, SVS~13 defies a clear classification as
either FUor or EXor. While FUors generally show late-type low gravity absorption spectra
in the infrared \citep{Greene2008} indicative of a luminous, optically dense disk,  
the prototypical EXor EX Lupi shows the
CO bandheads and many other optical lines in emission during extreme EXor outbursts \citep{Aspin2010}.
During minor outbursts, optical emission lines are observed, but CO may be in photospheric absorption \citep{Herbig2001}. 
SVS~13 shows the CO bandheads in emission \citep{Takami2006}. 
In the context of this paper, the important conclusion is that
the elevated post-outburst level of accretion and therefore jet activity in SVS~13
has persisted for the past 2 decades to the present, so that the light curve of 
SVS~13 shares the long duration maximum with FUors, while spectroscopically, it resembles
an EXor, and the small outburst amplitude resembles neither. 
This suggests that at least for the younger, more deeply embedded accretion instability events,
the traditional two classes may not be appropriate, and that a continous range of outburst
characteristics may be a better way to understand this phenomenon.
The light curve also implies that the next outburst of SVS~13, if indeed these outbursts occur
repetitively about every 30 years, will start from a brighter state of SVS~13 than the previous one.
This would mean that in addition to the repetitive outbursts, we are also observing cumulative
changes in SVS~13.

\subsection{The [FeII] Microjet}
The seeing-limited integral field spectroscopy of \citet{Davis2011} had only
marginally resolved the [FeII] emission around SVS~13. Our adaptive-optics corrected
OSIRIS data show that [FeII] traces a high-velocity microjet that extends from
the source SVS~13 into the area of the most recent molecular hydrogen bubble.
The results are summarized in the false-color images (Fig.~3). Here, the wavelength-integrated
continuum-subtracted flux of the H$_2$~1--0~S(1) line was Lucy-Richardson deconvolved and is displayed in the 
red channel. To show the location of the stellar central object in SVS~13 without introducing
the artifacts from imperfect deconvolution of a dominant bright source, we have added
10\% of the deconvolved flux of two continuum wavelength channels to the continuum-subtracted image.
This produces, in effect, a high-dynamic-range version of the SVS~13 image for the purpose of
showing the relationship of features at different wavelengths. Note that the wavelength
channels used here were different from those used as the PSF. The resulting deconvolved
continuum image is therefore not the trivial solution of deconvolving the PSF with itself.
The green channel of Fig. 3 shows the deconvolved integral over the [FeII] line at 1.644 $\mu$m.
In the same way as described above, a fraction of the continuum wavelengths immediately
adjacent to the line was added to the data to mark the location of the stellar source.
Finally, to adjust the color balance of the stellar object to white, the same fraction of
the continuum wavelength channels on either side of the [FeII] line was assigned to the
blue channel.

Our Lucy-Richardson deconvolved images of the high-excitation [FeII] shocks associated with
the jets show that the jet is very narrow, about 20 - 40~mas wide. 
The velocity diagram in Fig.~9 shows that the [FeII] emission of the jet is blueshifted 
by -140 to -150 km~s$^{-1}$ relative to the systemic velocity
of the molecular material around SVS~13 \citep{Warin1996}. 
A comparison of Figs. 7 and 9 demonstrates that the [FeII] emission is more blueshifted than the
H$_2$ S(1) emission. 
The direct superposition of the H$_2$ and [FeII] emission
line images shows that the bright portions of the microjet [FeII] emission extend up to the rim of the H$_2$ bubble,
and that faint traces of [FeII] emission can be detected up to about the center of the bubble.
The jet [FeII] intensity drops by a factor of $\approx$~20 at the
bubble surface. In fact, the brightest [FeII] along the jet axis is seen directly upwind
from the H$_2$ bubble surface. The total 1.644~$\mu$m [FeII] flux is dominated by the emission outside
of the expanding H$_2$ bubble. Since the front edge of the bright jet component near the bubble rim
is not sharply defined, and subject to the degree of Lucy-Richardson deconvolution and the
different quality of the adaptive optics correction achieved at the two epochs, we could not directly
measure the proper motion of the jet front edge. From Fig.~3 it seems clear that the [FeII] shock emission from the jet
is surrounded by an envelope of entrained ambient material radiating in the low-excitation shocked H$_2$ lines,
and that at least the bright portions of the jet [FeII] emission terminate at the bubble rim. From the proper
motion of the bubble front side, and the rate of bubble expansion (section 4.2), the expected, but not
directly measured, proper motion of  
the side of the bubble facing SVS~13 (the back side in the direction of motion) 
is 6.6~mas~yr$^{-1}$, which we
also take as the proper motion of the front end of the bright jet. 
The length of the [FeII] jet is therefore changing
only very slowly, at a rate of $\approx$ 7 mas year$^{-1}$, and the centroid of the [FeII] flux 
therefore moves at less than half this speed. This is responsible for the impression noted
already by \citet{Davis2006} that the [FeII] emission looked stationary.
What we see from the jet in the [FeII] emission line are either internal shocks or the
shocks resulting from interaction with ambient molecular material that then gets entrained
by the jet. The bulk of the jet material propagates into the area of the H$_2$ bubble,
but the excitation conditions are, apparently, less favorable to the formation of such shocks
radiating in [FeII]. 

\begin{figure}
\begin{center}
\figurenum{9}
\includegraphics[scale=0.7,angle=0]{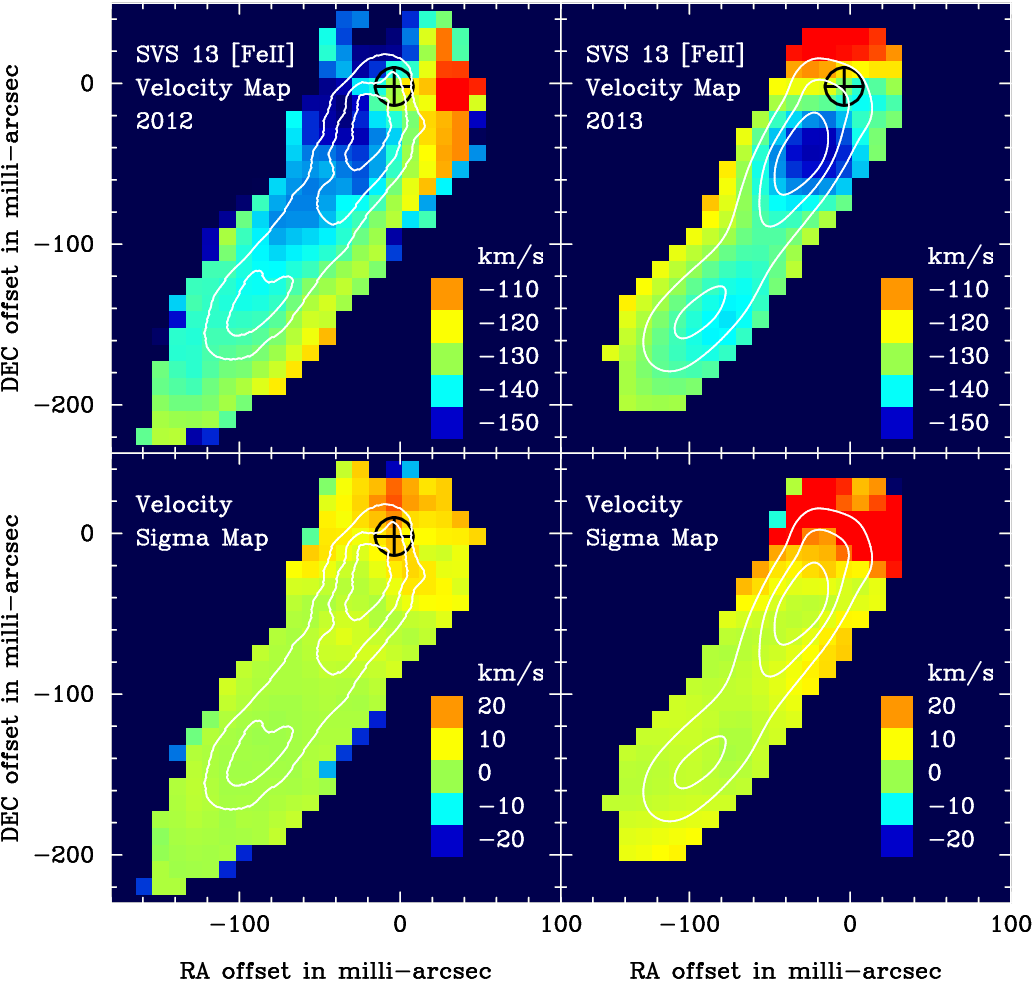}
\caption{
Top panels: Velocity map of the continuum-subtracted [FeII] 1.644 $\mu$m wavelength data cubes 
of the SVS~13 jet, deconvolved using the Lucy-Richardson algorithm, for the 2012 and 2013 data, respectively. 
The bottom panels show the corresponding velocity sigma maps.
The white contours outline the line integrated [FeII] flux distribution.
It clearly shows that all the line emission
from the jet is blue-shifted relative to the systemic velocity of SVS~13.
A color version of this figure is available in the electronic version of this paper.
}
\end{center}
\end{figure}

We have tried, both on the 2012 and 2013 data, to detect the signature of jet
rotation. Initially, using only the 2012 data, such rotation appeared to be indicated \citep{Hodapp2013}.
However, the 2013 data, and the re-reduced 2012 data with the new calibrations, 
did not confirm this (Fig. 9). The velocity pattern measured in 2013
was more confused and, if anything, a faint indication of the opposite rotation direction
was found. 
In Fig.~9 (lower panels) we show the rms variations of the velocity measurements
on the individual data cubes that were coadded to form the velocity maps in the 
top panels. These indicate that the errors of the velocity maps are below 10~km~s$^{-1}$
in most parts of the [FeII] jet.
We conclude that the excellent spatial resolution and moderate spectral 
resolution of OSIRIS are not sufficient to resolve the kinematic signature of jet
rotation in SVS~13.

\subsection{Atomic Hydrogen Emission}

The permitted atomic hydrogen line emission from SVS~13 was found 
to be broad with line widths of $\approx$ 180~$\pm$~10~km~s$^{-1}$
for Br$\gamma$ by \citet{Davis2001} 
and for Br-12 by \citet{Davis2003}. 
The emission is
centered on the position of the continuum source, but spatially unresolved.
Our data cube in the Hn3 filter contains the Br-12 and Br-13 hydrogen recombination lines that 
trace the same hot hydrogen recombination regions as the more frequently used Br-$\gamma$ does.
In Fig. 10, we show the deconvolved images of SVS~13 across the Br-13 emission line
after subtraction of the continuum emission. 
This figure confirms that emission in the atomic hydrogen
recombination lines is spatially centered on the young star, and has the same flux profile as the continuum,
i.e. we see no indication that the Br-13 emission is extended. Figure 10 has 10 mas pixels (2.35~AU), and any
systematic differences between the line and continuum PSF are well below that angular scale. We conclude
that the zone of atomic hydrogen emission, presumably the accretion disk itself, is less that 2 AU
in extent. It is actually expected to be only of order of the dimensions of the star itself, i.e.,
about 2 orders of magnitude smaller than this limit.
Our velocity data are consistent with the higher
spectral resolution data of \citet{Davis2001} who found that the Br$\gamma$ line is centered at -25~($\pm15$)~km~s$^{-1}$
relative to the systemic velocity of the SVS~13 core, essentially at the same velocity as the star. 

\clearpage
\begin{figure}
\begin{center}
\figurenum{10}
\includegraphics[scale=1.2,angle=0]{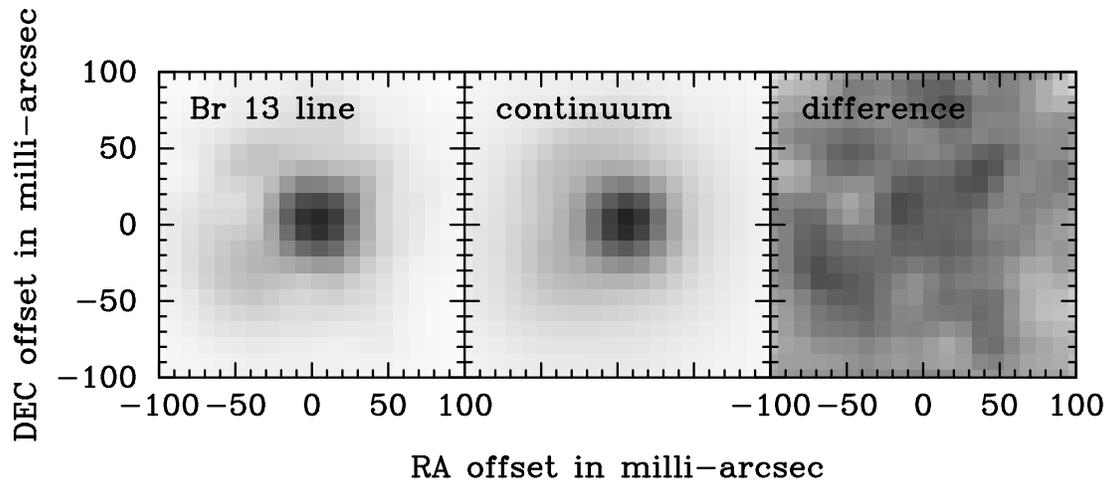}
\caption{
Left: Continuum-subtracted image of SVS~13 in the Br~13-4 line of atomic Hydrogen.
Middle: Continuum adjacent to the line.
Right: Ratio of these two, showing that the PSF is the same for the line emission and
the continuum, within the limitations of the noise.
This demonstrates that the H Br 13 line emission remains unresolved at the level of 
about 50~mas ($\approx$~2~AU).
}
\end{center}
\end{figure}

\clearpage
\section{CONCLUSIONS}
We have presented adaptive optics corrected integral field spectroscopy
of the young outflow source SVS~13 in NGC~1333. The H$_2$ 1--0 S(1) line at
2.122 $\mu$m, indicating low-velocity shocks, the higher excitation [FeII] line
of the micro-jet at 1.644 $\mu$m, and atomic hydrogen emission in the HI12-4 and HI13-4
lines were analyzed and lead us to the following main conclusions:
\begin{enumerate}
\item The HH~7-11 outflow originates in SVS~13, which is identical to VLA~4B.
\item The outflow, at present, originates as a micro-jet of $\approx$~0$\farcs$2 length,
detectable in [FeII] and oriented at P.A. 145$^{\circ}$.
\item The formation of the youngest partly formed bubble visible in H$_2$ emission 
can be traced back to the $\approx$~1990 outburst.
\item The bright parts of the [FeII] microjet reach up to the boundary of the
H$_2$ S(1) bubble, but fainter [FeII] emission can be traced another $\approx$~0$\farcs$2
to near the center of the bubble.
\item Beyond that, H$_2$ S(1) emission outlines a curved path of the jet.
\item The orientation of the next two bubbles at P.A.~$\approx$~159$^{\circ}$
is roughly point-symmetric to the orientation of outflowing material
in the counter-jet found by \citet{Bachiller2000} in CO and \citet{Raga2013} at 4.5~$\mu$m.
\item The chain of bubble fragments and their proper motions suggest that bubble-generating 
events are occurring repetitively, roughly every 30 years, at the present time.
\item The formation of a series of expanding bubbles within the outflow by a series of eruptive events provides
an explanation for the widening of the outflow cavity.
\item Atomic hydrogen emission in the HI12-4 and HI13-4 (Brackett series) lines is 
detected around the continuum position of SVS~13, indicating ongoing accretion onto
the star.
\item The outflow source SVS~13 remains at or near the peak brightness reached
during the 1990 outburst. The light curve therefore resembles
that of FUor type objects, while the emission line spectrum matches the characteristics
of EXors. SVS~13 therefore represents an object somewhere between those classical classes.
\end{enumerate}

\acknowledgments

Most of the data presented herein were obtained at the W.M. Keck Observatory,   
which is operated as a scientific partnership among the 
California Institute of Technology,
the University of California and NASA.
The Observatory was made possible by the generous financial support of the W.M. Keck Foundation.
Some photometric data on SVS~13 were obtained at the IRIS telescope on Cerro Armazones,
which is operated under a cooperative agreement between 
the "Astronomisches Institut, Ruhr Universit\"at Bochum", Germany, the "Universidad Catolica del Norte" in Antofagasta, Chile,
and the Institute for Astronomy, University of Hawaii, USA. Construction of the IRIS infrared
camera was supported by the National Science Foundation under grant AST07-04954.
The operation of the IRIS telescope is supported by the ``Nordrhein-Westf\"alische Akademie der
Wissenschaften und der K\"unste'' in the framework of the academy program by the Federal
Republic of Germany and the state of Nordrhein-Westfalen.
We wish to thank 
Angie Barr Dominguez,
Thomas Dembsky, 
Holger Drass, 
Lena Kaderhandt, 
Michael Ramolla 
and 
Christian Westhues
for
operating the IRIS telescope for the acquisition of the data used in this paper,
Ramon Watermann for writing the data reduction pipeline,
and Roland Lemke for technical support.
We thank Tuan Do and Jessica Lu for kindly providing the information about the
OSIRIS spaxel scales on the Keck I and Keck II telescopes.

This publication makes use of data products from the Two Micron All Sky Survey, which is
a joint project of the University of Massachusetts and the Infrared Processing and Analysis Center/
California Institute of Technology, funded by the National Aeronautics and Space Administration
and the National Science Foundation.
This publication also uses archival data obtained with the Spitzer Space Telescope, which is operated
by the Jet Propulsion Laboratory, California Institute of Technology under a contract with NASA.

\clearpage


\end{document}